# Multimodal Imaging-based Material Mass Density Estimation for Proton Therapy Using Physics-Constrained Deep Learning

Chih-Wei Chang[1], Raanan Marants[2], Yuan Gao[1], Matthew Goette[1], Jessica E. Scholey[3],

Jeffrey D. Bradley[1], Tian Liu[1], Jun Zhou[1], Atchar Sudhyadhom[2*] and Xiaofeng Yang[1*]

[1]Department of Radiation Oncology and Winship Cancer Institute, Emory University, Atlanta, GA 30308

[2]Department of Radiation Oncology, Brigham & Women's Hospital/Dana-Farber Cancer Institute/Harvard Medical School, Boston, MA 02115

[3]Department of Radiation Oncology, The University of California, San Francisco, CA 94143

*Corresponding author

Email: xiaofeng.yang@emory.edu and atchar_sudhyadhom@dfci.harvard.edu

**Running title:** MRI/DECT Material Mass Density Mapping

## Abstract

Mapping computed tomography (CT) number to material property dominates the proton range uncertainty. This work aims to develop a physics-constrained deep learning-based multimodal imaging (PDMI) framework to integrate physics, deep learning, magnetic resonance imaging (MRI), and advanced dual-energy CT (DECT) to derive accurate patient mass density maps. Seven tissue substitute MRI phantoms were used for PDMI-based material calibration. The training inputs are from MRI and twin-beam dual-energy images acquired at 120 kVp with gold and tin filters. The feasibility investigation included an empirical DECT correlation and four residual networks (ResNet) derived from different training inputs and strategies by the PDMI framework. PRN-MR-DE and RN-MR-DE denote ResNet trained with and without a physics constraint using MRI and DECT images. PRN-DE and RN-DE represent ResNet trained with and without a physics constraint using DECT-only images. For the tissue surrogate study, PRN-MR-DE, PRN-DE, and RN-MR-DE result in mean mass density errors: -0.72%, 2.62%, -3.58% for adipose; -0.03%, -0.61%, and -0.18% for muscle; -0.58%, -1.36%, and -4.86% for 45% HA bone. The retrospective patient study indicated that PRN-MR-DE predicted the densities of soft tissue and bone within expected intervals based on the literature survey, while PRN-DE generated large density deviations. The proposed PDMI framework can generate accurate mass density maps using MRI and DECT images. The physics-constrained training can further enhance model efficacy, making PRN-MR-DE outperform RN-MR-DE. The patient investigation also shows that the PDMI framework has the potential to improve proton range uncertainty with accurate patient mass density maps.

# 1 Introduction

Compared to traditional photon radiation therapy, proton therapy has demonstrated a reduction in patients' side effects and unplanned hospitalizations while achieving comparable tumor control (Baumann *et al.*, 2020). Protons offer the physical advantage of stopping just after depositing highest dose, sparing healthy tissues beyond the tumor volume (Goitein, 1985; Dinges *et al.*, 2015). One of the critical factors that dominates the quality of proton treatment planning is the accuracy of the dose calculation algorithm. Dose calculation algorithms, which may be analytical (Schuemann *et al.*, 2015; Yepes *et al.*, 2018; Liang *et al.*, 2019) or Monte Carlo-based (Paganetti *et al.*, 2008; Saini *et al.*, 2017; Chang *et al.*, 2020), require the relative stopping power (RSP) or mass density of materials to simulate transport phenomena within patients. Conventional treatment planning systems use computed tomography (CT) numbers to material property curves calibrated by the stoichiometric method (Schneider *et al.*, 1996; Schneider *et al.*, 2000) to convert Hounsfield units (HU) to material RSP or mass densities from CT images. However, this approach is based on fitting data to a linear model which is limited in how mass densities are categorized for different tissue types (Ratner, 2011; Wohlfahrt *et al.*, 2020), and materials with different compositions may result in the same attenuation measured by CT scanners. To account for this uncertainty in HU-to-RSP or HU-to-mass-density conversion, a 2.5-3.5% proton range margin (Paganetti, 2012) is usually adapted for treatment planning.

Dual-energy computed tomography (DECT) can effectively characterize tissues and detect lesions (Patel *et al.*, 2013; McCollough *et al.*, 2015). The parametric maps inferred from DECT images, such as the effective atomic number and relative electron density, can be used to derive proton RSP for analytical dose calculation (Bär *et al.*, 2017; Wohlfahrt *et al.*, 2017). Virtual monochromatic images (VMI) derived from DECT can also increase image quality and reduce beam-hardening artifacts compared to CT images from a polychromatic spectrum (Yu *et al.*, 2011; Yu *et al.*, 2012; Grant *et al.*, 2014). In more recent work by Medrano *et al. (Medrano et al., 2022)*, they developed an RSP prediction method based on the sinogram domain that achieved bony-tissue uncertainty of 0.8%. In general, accurate proton RSP maps can be acquired from DECT images (Yang *et al.*, 2010), and this approach leads to more exact proton dose calculation than the method using conventional CT imaging (Zhu and Penfold, 2016; Bär *et al.*, 2018). Wohlfahrt *et al. (Wohlfahrt et al., 2019)* recommended patient-specific DECT-based RSP prediction for treatment planning. They concluded that intra-patient adipose and soft tissue diversity were 5.6% and 9.8%, and the mean RSP deviation was 1.2% between the Hounsfield look-up table and patient-specific RSP map.

Meanwhile, magnetic resonance imaging (MRI) demonstrates favorable soft tissue contrast relative to CT and its utility in radiation oncology has seen a significant increase over the past decade. Synthetic CT-based MRI-only proton therapy has been investigated, and while others have concluded that an MRI-only treatment planning is feasible, but inaccurate bone delineation has the potential to lead to significant uncertainty (Rank *et al.*, 2013; Edmund *et al.*, 2014; Koivula *et al.*, 2016). Sudhyadhom (Sudhyadhom, 2017) proposed a method to directly determine material mean ionization potential ($I$) from MRI. Using this method, Scholey *et al.* (Scholey *et al.*, 2021) predicted RSP by using the Bethe-Bloch formula with material $I$ and relative electron densities (based on the CT stoichiometric method) from MRI and CT separately, and achieved RSP uncertainty within 1% for soft tissues. It would be of great interest to develop a model that can directly assimilate MRI and DECT information to generate material properties.

However, a fixed form of the linear model is used for the DECT-based stoichiometric method (Bourque *et al.*, 2014; Xie *et al.*, 2018) to fit data, and it is nontrivial for the model to utilize MRI directly. It typically requires extensive time and effort to gain the mechanistic understanding necessary to develop a new model that can simultaneously assimilate MRI and DECT data. Alternatively, modern machine learning (ML) methods can deploy models with flexible forms to effectively discover the underlying correlations between MRI/CT images and material characteristics. Su *et al.* (Su *et al.*, 2018) demonstrated that ML models could generate RSP mapping with the uncertainty of ~3% for cortical bone from DECT images. In Scholey *et al.* (Scholey *et al.*, 2022), they show that ML from low energy CT to higher energy CT (kVCT to MVCT) may be useful in the accurate generation of RSP maps with error reductions primarily due to improvements in electron density. RSP accuracy in that study was limited by a lack of compositional (elemental or molecular) information as available in DECT or MRI.

Many conventional ML methods favor the principle of Occam's razor: simplicity leads to the optimal model because of its generalizability and interpretability (Blumer *et al.*, 1987; Domingos, 1999). However, the strategy may not be applicable when dealing with a substantial amount of data (Champion *et al.*, 2019). In contrast, neural networks have been proven universal approximators (Hornik *et al.*, 1989), and modern deep learning (DL) models with hierarchical structures can learn complex patterns from data (LeCun *et al.*, 2015). Furthermore, physics-informed deep learning (Chang and Dinh, 2019; Karniadakis *et al.*, 2021; Chang *et al.*, 2022) integrates physics insights that could increase the accuracy and robustness of DL models, especially when dealing with noise or limited data.

In the present work, we use a physics model to regularize DL models during training and develop a physics-constrained DL-based multimodal imaging (PDMI) framework to predict patient mass density maps from MRI and DECT images. The framework can potentially adopt ML and DL models based on different applications. To evaluate the feasibility of the proposed method for clinical applications, we explore conditions by which DL models can benefit from the PDMI framework to infer material mass density maps accurately and effectively from MRI and DECT images.

## 2 Materials and methods

### 2.1 Tissue substitute phantoms and data acquisition

Five tissue substitute phantoms were created following the recipe by Scholey *et al. (Scholey et al., 2021)*, including adipose, muscle, skin, and spongiosa. We also added 45% HA bone as that's a typical composition of "standard" bone (whereas spongiosa is less dense and contains fat). The phantoms were made by homogeneous mixtures of deionized water, gelatin from porcine skin (surrogate of protein), porcine lard (surrogate of fat), and hydroxyapatite (HA). A small amount of sodium dodecyl sulfate (SDS) was added to the phantoms with water and lard to promote mixing and enhance homogeneity. Scholey *et al. (Scholey et al., 2021)* investigated each mixture to ensure that their physical characteristics were similar to actual biological tissues for the respective modality. Table 1 gives mass percent compositions of each mixture for tissue surrogates.

**Table 1.** Mass percent compositions of mixtures for tissue substitute phantoms.

| Tissue surrogate | Water | Gelatin (Protein) | Lard (Fat) | Hydroxyapatite | SDS |
|---|---|---|---|---|---|
| Adipose | | | 100 | | |
| Muscle | 74.78 | 19.97 | 5.0 | | 0.25 |
| Skin | 75.0 | 25.0 | | | |
| Spongiosa | 26.61 | 11.83 | 47.43 | 12.81 | 1.32 |
| 45% HA Bone | 55.0 | | | 45.0 | |

Two additional fresh animal tissue phantoms were made from *ex vivo* porcine blood with either porcine brain (brain phantom) or porcine liver (liver phantom). To determine the elemental compositions of these phantoms, multiple small brain and liver samples were weighed, dehydrated, crushed, and sent off for combustion analysis to determine their elemental compositions. These results, along with the known amount of blood (assumed to have the same elemental composition as water) and hydrated tissue comprising the brain and liver phantoms, were then used to compute the elemental compositions of the animal tissue phantoms. The mass densities of each tissue substitute and animal tissue phantom were measured by a high-precision scale (HK-3200A, Mars Scale Corporation, Canada) and volumetric pipettes. Table 2 summarizes each tissue surrogate's measured mass density and physical properties.

**Table 2.** Physically measured mass densities ($\rho_{meas}$), mean excitation energies ($I$), and elemental mass percent compositions for tissue substitute phantoms.

| Tissue surrogate | $\rho_{meas}$ (g/cm$^3$) | $I$ (eV) | H | C | N | O | Na | P | S | Ca |
|---|---|---|---|---|---|---|---|---|---|---|
| Adipose | 0.936 | 61.9 | 12.37 | 77.70 | 0.16 | 9.77 | | | | |
| Brain | 1.028 | 77.3 | 10.96 | 6.02 | 0.73 | 82.24 | | | 0.05 | |
| Muscle | 1.064 | 76.7 | 10.37 | 12.44 | 3.06 | 73.86 | 0.02 | | 0.25 | |
| Liver | 1.065 | 77.2 | 10.60 | 8.98 | 2.05 | 78.25 | | | 0.12 | |
| Skin | 1.077 | 77.5 | 10.09 | 10.55 | 3.82 | 75.26 | | | 0.28 | |
| Spongiosa | 1.091 | 76.1 | 9.79 | 42.50 | 1.88 | 37.96 | 0.11 | 2.37 | 0.28 | 5.11 |
| 45% HA Bone | 1.417 | 108.5 | 6.24 | | | 67.48 | | 8.32 | | 17.96 |

Tissue substitute phantoms were scanned with a Siemens SOMATOM Definition Edge scanner using a head-and-neck (HN) TwinBeam dual-energy (TBDE) protocol with a CT dose index (CTDI$_{vol, 32cm}$) of 8.6 mGy and effective milliampere-seconds (mAs$_{eff}$) of 400. TBDE protocols acquire images at 120 kVp with gold (Au) and tin (Sn) filters, and we refer to images from low-energy and high-energy spectra as DECT HighE and DECT LowE images. DECT parametric maps were generated from Siemens Syngo.Via including relative electron density and VMI of 80 keV. A Siemens MAGNETOM Aera 1.5T scanner was used to acquire MR images of the phantoms using a 3D T1-weighted Dixon VIBE sequence, generating water-only (T1-DW) and fat-only (T1-DF) images, and a 3D T2-weighted short tau inversion recovery SPACE sequence (T2-STIR).

Table 3 summarizes the acquisition parameters used to acquire DECT and MR images. Figure 1 depicts the axial view of phantom images acquired from MRI and DECT scans. The dimension of each phantom was 5.7x5.7x12.9 cm$^3$, and each phantom was placed in a cylindrical water container with a diameter and height equal to 16.83 cm and 25.4 cm. MRI T1 and T2 scans contain 208x288x176 and 256x256x208 voxels, and DECT scans include 512x512x545 voxels.

**Table 3.** DECT and MRI acquisition parameters.

| | Scanner | Siemens SOMATOM Definition Edge |
|---|---|---|
| | Collimation | 64x0.6 mm |
| | Voxel size | 0.977x0.977x0.5 mm$^3$ |
| | Field of view | 500 mm |
| | X-ray tube voltage | 120 kVp with Au and Sn filters |
| | Scanner | Siemens MAGNETOM Aera 1.5T |
| | 3D DIXON VIBE voxel size | 1.247x1.247x1.2 mm$^3$ |
| T1 | Repetition time (TR) | 7.76 ms |
| | Echo time (TE) | 2.39 ms |
| | 3D STIR SPACE voxel size | 1.016x1.016x1.1 mm$^3$ |
| T2 | Repetition time (TR) | 3500 ms |
| | Echo time (TE) | 248 ms |

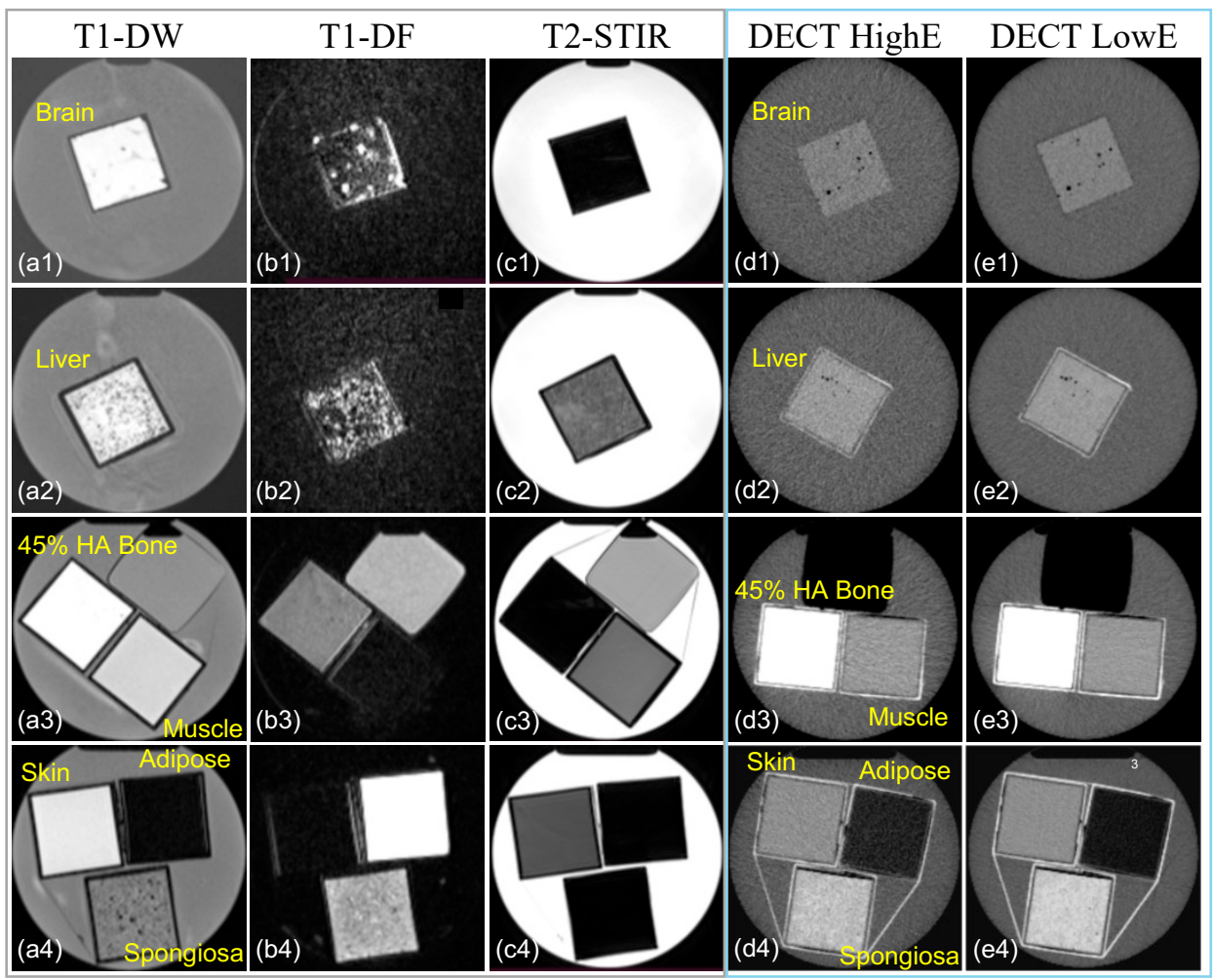

|  | T1-DW | T1-DF | T2-STIR | DECT HighE | DECT LowE |

**Figure 1.** Transversal representation of tissue substitute phantom images acquired from (a1)-(a4) T1-weighted Dixon water-only (T1-DW) MRI, (b1)-(b4) T1-weighted Dixon fat-only (T1-DF) MRI, (c1)-(c4) T2-weighted short tau inversion recovery (T2-STIR), (d1)-(d4) high-energy spectrum imaging of twin-beam dual-energy CT (DECT HighE), and low-energy spectrum imaging of twin-beam dual-energy CT (DECT LowE).

A retrospective head-and-neck (HN) patient study was conducted for a proof-of-concept test of the proposed framework due to the lack of ground truth mass density maps. The HN patient was scanned with identical MRI and DECT imaging protocols as the phantom study. Since patient images were heterogeneous, MRI images were first rigidly registered to DECT images, and then deformable image registration was applied using Velocity[TM]. Ultimately, MRI images were resampled to match the resolution of DECT images, which includes 512x512x470 voxels with a voxel size of 0.977x0.977x1 mm[3].

## 2.2 Physics-constrained deep learning-based multimodal imaging framework

Figure 2 shows the physics-constrained DL-based multimodal imaging (PDMI) framework to generate material mass density maps from MRI and DECT images and validate the phantom results through a proton experiment. The model inputs from MRI include T1-DF, T1-DW, and T2-STIR, while the inputs from DECT images are HighE, LowE, $\rho_e$, and 80-keV VMI. The framework can adapt arbitrary DL models and train DL models with a combination of MRI and DECT images or DECT-only images. The upper part of Figure 2 illustrates the training workflow that involves two branches to train DL models with different loss functions. Figure 2(a32) shows a conventional mass density loss function ($\mathcal{L}_\rho$) defined by mean square error with ground truth mass density ($\rho_{meas}$) from the physical measurement of mass density for tissue substitute phantoms. Unlike conventional DL training, Figure 2(a33) shows the proposed physics-constrained loss function ($\mathcal{L}_{physics}$) using physics insights without requiring ground truth mass density, which requires separate measurement that can introduce additional uncertainty. The physics insights can be quantified from a well-established physics-based models to constrained DL models. Section 2.2.2. describes how we tightly integrate physics insights (Eq. (4)) into DL training in this work. Figure 2(a43) shows the experiment setup for proton measurement (see Section 2.2.3) that can be used to validate the accuracy of DL models. The lower part of Figure 2 depicts the workflow for patient applications. Patients' DECT and MIR images will be input to a pre-trained and validated ResNet model. Then the ResNet will generate the corresponding mass density map to support potential proton Monte Carlo dose calculation for treatment planning.

### 2.2.1 Deep learning model and training data

We used PyTorch (Paszke *et al.*, 2019) to implement residual networks (ResNet) in Figure 2(a31) from our previous work (Chang *et al.*, 2022) due to its accuracy in learning from data through hierarchical modelling (He *et al.*, 2016). The ResNet networks trained by the physics-constrained loss and conventional loss are named physics-constrained ResNet (PRN) and conventional ResNet (RN). We also add a suffix, "-MR-DE" or "-DE," to PRN and RN to indicate whether the model is trained with MRI and DECT images or DECT-only images. For instance, PRN-MR-DE presents the ResNet trained with the physics-constrained loss using MRI and DECT images. ReLU(Nair and Hinton, 2010) is the activation function used in the network, and the details of model structures are given in Appendix A.

Circular volumes of interest (VOI) were manually contoured for each tissue surrogate with diameters of 40 and 28 pixels for MRI and DECT images. Data from each VOI were subsequently arranged in a 1-dimensional (1D) array to form a DL input matrix with the size of $n_{vox} \times n_{inp}$ where $n_{vox}$ and $n_{inp}$ denote the total voxels within VOI from each material and input images such as T1-DF, T1-DW, T2-STIR, HighE, LowE, $\rho_e$, and 80-keV VMI for PRN-MR-DE. The $n_{vox}$ contained 444,465 and 226,590 voxels for training and testing from each VOI on MRI and DECT images for seven materials.

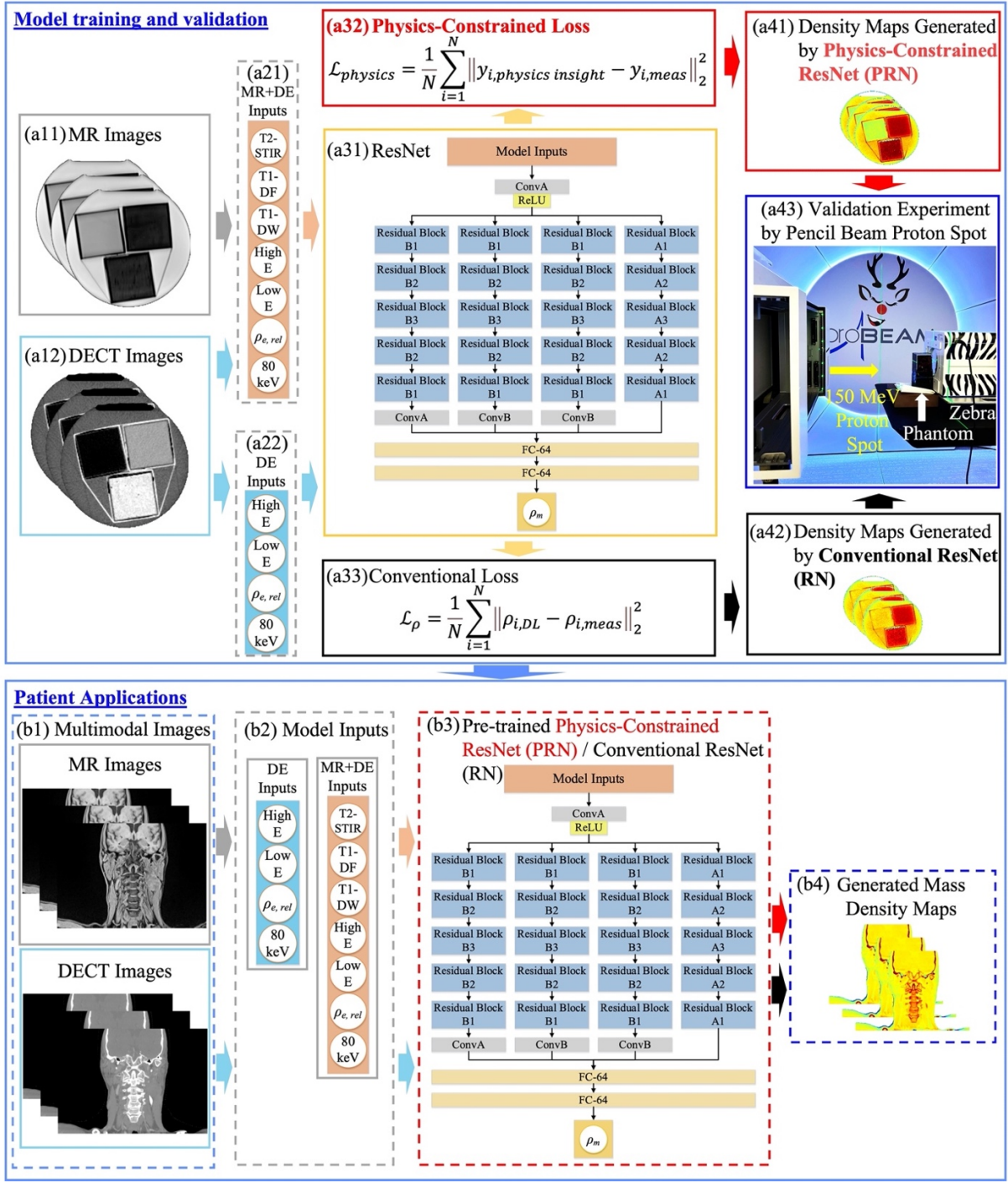

**Figure 2.** Physics-constrained deep learning-based multimodal imaging (PDMI) framework for material mass density inference from (a11) MRI and (a12) DECT images. (a31) DL models adapted in the framework. ResNet was implemented where ConvA, ConB, and Residual Blocks represent different convolutional components. (a32) Conventional mass density loss and (a33) physics-constrained loss for training. Material mass densities generated by (a41) conventional ResNet (RN) and (a42) physics-constrained ResNet (PRN). (a43) Validation experiment to obtain measured RSP for tissue substitute phantoms using a 150-MeV proton spot and Zebra (IBA Dosimetry, Germany). (b1-b4) Inference of mass density map for patient applications using MRI and DECT images as inputs for PRN and RN.

### 2.2.2 Mass density and physics-constrained loss functions

The PDMI framework relies on supervised learning to train DL models that can infer material mass densities from MRI and DECT images. This inverse modelling is ill-posed, and it can be solved by optimizing the loss function defined by Eq. (1), which includes conventional mass density ($\mathcal{L}_\rho$) and physics-constrained ($\mathcal{L}_{physics}$) losses. For $\delta$=1, physics-constrained training is performed, and conventional training is used when $\delta$=0.

$$\mathcal{L} = (1 - \delta) \times \mathcal{L}_\rho + \delta \times \mathcal{L}_{physics} \tag{1}$$

Eq. (2) defines a conventional mass loss function that requires measured material mass density ($\rho_{meas}$) as the targets. The $\rho_{DL}$, $N$, and $i$ denote the predicted mass density from DL models, total number of voxels, and i$^{th}$ voxel.

$$\mathcal{L}_\rho = \frac{1}{N} \sum_{i=1}^{N} \left\| \rho_{i,DL} - \rho_{i,meas} \right\|_2^2 \tag{2}$$

Meanwhile, the PDMI framework also adapts physics insights to form physics-constrained loss functions given by Eq. (3), where $y_{physics\ insight}$ and $y_{meas}$ denote the physics insights and corresponding measured quantities of the physics insights, respectively. We select $y_{physics\ insight}$ as the empirical HU model from the stoichiometric calibration method(Rutherford *et al.*, 1976; Schneider *et al.*, 1996; Bourque *et al.*, 2014), and $y_{meas}$ becomes the CT numbers in this study.

$$\mathcal{L}_{physics} = \frac{1}{N} \sum_{i=1}^{N} \left\| y_{i,physics\ insight} - y_{i,meas} \right\|_2^2 \tag{3}$$

Eq. (4) gives the empirical HU model with the assumption that Hounsfield unit ($HU$) = $1000(\mu/\mu_w-1)$ where $\mu$ and $\mu_w$ are linear attenuation coefficients of the material and water. The $k_{ph}$, $k_{coh}$, and $k_{incoh}$ are energy-dependent coefficients associated with the photon-electric, coherent, incoherent effects, respectively. These constants can be determined from the stoichiometric calibration method(Schneider *et al.*, 1996; Schneider *et al.*, 2000) with least-square fitting. In this work, the estimated $k_{ph}$, $k_{coh}$, and $k_{incoh}$ coefficients are 9.094 x $10^{-6}$, 1.064 x $10^{-3}$, and 5.988 x $10^{-1}$.

$$y_{physics\ insight} \equiv HU_{80keV} = 1000 \left[ \tilde{\rho} \frac{k_{ph} \tilde{z}^{3.62} + k_{coh} \tilde{z}^{1.86} + k_{incoh}}{k_{ph} \tilde{z}_w^{3.62} + k_{coh} \tilde{z}_w^{1.86} + k_{incoh}} - 1 \right] \tag{4}$$

By defining $\tilde{z} \equiv z_{3.62}$ and $\hat{z} \equiv z_{1.86}$, the material $z_n$ values can be derived by Eq. (5)(Mayneord, 1937; Spiers, 1946) where n is 3.62 or 1.86. The $i$, $\omega$, $Z$, and $A$ denote $i^{th}$ element, weight fraction, atomic number, and atomic mass number. The $z_{3.62}$ and $z_{1.86}$ are 7.522 and 7.115 for water.

$$z_n = \left( \sum_i \frac{\omega_i \frac{Z_i}{A_i}}{\sum_i \omega_i \frac{Z_i}{A_i}} Z_i^n \right)^{\frac{1}{n}} \tag{5}$$

The relative electron density derived using the DL model output ($\tilde{\rho}$) is defined by Eq. (6) with the electron density of water ($\rho_{e,w}$) equal to 3.343 x $10^{23}$ e⁻/cm³.

$$\tilde{\rho} \equiv \frac{\rho_{DL} \sum_i \omega_i \frac{Z_i}{A_i}}{\rho_{e,w}} \tag{6}$$

The value of $\rho_{DL}$ is queried from DL models during each training iteration that allows Eq. (4) to be updated and the physics loss given by Eq. (3). The target is 80-keV VMI for the physics-constrained training because the image provides the optimal image-noise and beam-hardening ratios (Yu *et al.*, 2012; Wohlfahrt *et al.*, 2017; Wang *et al.*, 2019).

### 2.2.3    Proton RSP measurement for tissue substitute phantoms

A 150 MeV proton spot delivered by Varian ProBeam System (Varian Medical Systems, Palo Alto) was used to determine the measured RSP of seven tissue substitute phantoms including adipose, brain, muscle, liver, skin, spongiosa, and 45% HA bone. The dimension of each phantom container was 5.7(width)x5.7(height)x12.9(length) cm³. Figure 2(a43) illustrates the setup for the experiment with Zebra (IBA Dosimetry, Germany) to measure proton 80% distal range (R80). The water equivalent thickness (WET) was computed by taking differences between R80 measurements with and without phantoms. WET values were divided by the physical width of phantom containers to obtain measured RSP for each material given in Table 2.

The Bethe-Bloch equation (Bichsel, 1969) can be rearranged as Eq. (7) to obtain the reference mass density ($\rho_{ref}$) using measured RSP ($RSP_{meas}$) where $m_e$, $c$, $\beta$, and $I$ are the electron mass, speed of light, proton (150 MeV) velocity relative to the speed of light, and mean ionization potential of materials. The mean ionization potential of water ($I_{water}$) is 75 eV (ICRU37, 1987). Table 4 gives the measured RSP, reference mass densities, and other material characteristics defined by Eq. (5) for each phantom.

$$\rho_{ref} = RSP_{meas} \left\{ \frac{\sum_i \omega_i \frac{Z_i}{A_i}}{\rho_{e,w}} \cdot \frac{ln\left[\frac{2m_e c^2 \beta^2}{I(1-\beta^2)}\right] - \beta^2}{ln\left[\frac{2m_e c^2 \beta^2}{I_{water}(1-\beta^2)}\right] - \beta^2} \right\}^{-1} \tag{7}$$

**Table 4.** Material characteristics for tissue substitute phantoms.

| Tissue surrogate | $RSP_{meas}$ | $\rho_{ref}$ (g/cm$^3$) | $\tilde{z}$ | $\hat{z}$ |
|---|---|---|---|---|
| Adipose | 0.979 | 0.947 | 5.902 | 5.502 |
| Brain | 1.014 | 1.020 | 7.445 | 7.023 |
| Muscle | 1.056 | 1.067 | 7.406 | 6.954 |
| Liver | 1.054 | 1.063 | 7.421 | 6.993 |
| Skin | 1.067 | 1.082 | 7.447 | 7.004 |
| Spongiosa | 1.076 | 1.093 | 9.711 | 7.673 |
| 45% HA Bone | 1.344 | 1.476 | 13.152 | 11.025 |

### 2.3 Empirical model for DECT parametric mapping

Eq. (8) shows the empirical model (Beaulieu *et al.*, 2012) to estimate material mass densities using DECT parametric maps of relative electron density ($\rho_e$) and effective atomic number ($Z_{eff}$) obtained from Siemens Syngo.Via. This linear fitting model is not valid for low-density material such as an inflated lung, so a constant value of 0.26 g/cm$^3$ is used when materials' relative electron densities are lower than 0.37. We implemented the empirical model in Matlab R2021b and compared it to ResNet models trained by the PDMI framework.

$$\rho(g/cm^3) = \begin{cases} -0.1746 + 1.176\rho_e & \rho_e \geq 0.37 \\ 0.26 & \rho_e < 0.37 \end{cases} \tag{8}$$

### 2.4 Evaluation

The mean percentage error (MPE) is the evaluation metric to quantify model accuracy for tissue surrogate phantom analyses with the reference mass densities from proton measurement. Eq. (9) defines MPE where $x$, *REF*, $i$, and $N$ denote the voxel quantity, reference values, i$^{th}$ voxel, and total voxels.

$$MPE = \frac{1}{N}\sum_{i=1}^{N}\left(\frac{x_i - x_{i,REF}}{x_{i,REF}}\right) \times 100\% \tag{9}$$

### 3 Results

The physics-constrained DL-based multimodal imaging (PDMI) framework allows ResNet to be trained with a physics-constrained loss or a conventional mass density loss. The investigated models include ResNet trained with a physics-constrained loss using MRI and DECT images or DECT-only images (PRN-MR-DE or PRN-DE) and ResNet trained with a conventional mass density loss using MRI and DECT images or DECT-only images (RN-MR-DE or RN-DE).

## 3.1 Inference of material mass densities for tissue substitute phantoms

MPE results for the empirical, conventional, and physics-informed models are shown in Table 5, with lowest MPE value for each phantom in bold. PRN-MR-DE yields the lowest MPE values for most phantom materials. The additional MRI data adapted by PRN-MR-DE improve the model accuracy for all phantoms compared to PRN-DE. For example, PRN-MR-DE improves the errors from PRN-DE by 1.9% and 1.08% for adipose and skin. In the case where RN-MR-DE results in considerable uncertainty, the physics-constrained training usually can regularize ResNet to deliver lower model errors. For instance, PRN-MR-DE reduces the errors from RN-MR-DE by 2.86% and 4.28% for adipose and 45% HA bone. The empirical model achieves the optimal error of -0.62% for adipose.

**Table 5.** Comparisons of mass densities between the reference and different models.

| Phantom | Mean Percentage Error (MPE) | | | | |
| | Empirical Model | Conventional | | Physics-Informed | |
| | | RN-DE | RN-MR-DE | PRN-DE | PRN-MR-DE |
| --- | --- | --- | --- | --- | --- |
| Adipose | **-0.62%** | 7.65% | -3.58% | 2.62% | -0.72% |
| Brain | -0.37% | 0.37% | 0.49% | 0.34% | **-0.29%** |
| Muscle | 0.09% | -0.93% | -0.18% | -0.61% | **-0.03%** |
| Liver | 0.27% | -0.71% | 0.34% | -0.68% | **0.22%** |
| Skin | 0.56% | 0.51% | **-0.13%** | 1.24% | 0.16% |
| Spongiosa | 1.02% | 1.20% | **0.05%** | 3.83% | 3.13% |
| 45% HA Bone | -0.78% | -5.72% | -4.86% | -1.36% | **-0.58%** |

Figure 3 shows the boxplot of different models' mass density distributions for each phantom. RN-DE yields an unreasonable density range for adipose, since the upper bond reaches 1.2 g/cm$^3$. When the reference mass densities of adipose and 45% HA bone are not included in interquartile ranges by RN-MR-DE, the physics-constrained training can regularize the model responses. For instance, PRN-MR-DE predicts the median adipose and 45% HA bone values close to the reference. The empirical model delivers median values within interquartile ranges for most phantoms except skin and spongiosa. RN-MR-DE results in the optimal median value prediction for spongiosa. For most phantoms, the median values predicted by PRN-MR-DE are the closest to the reference compared to other models.

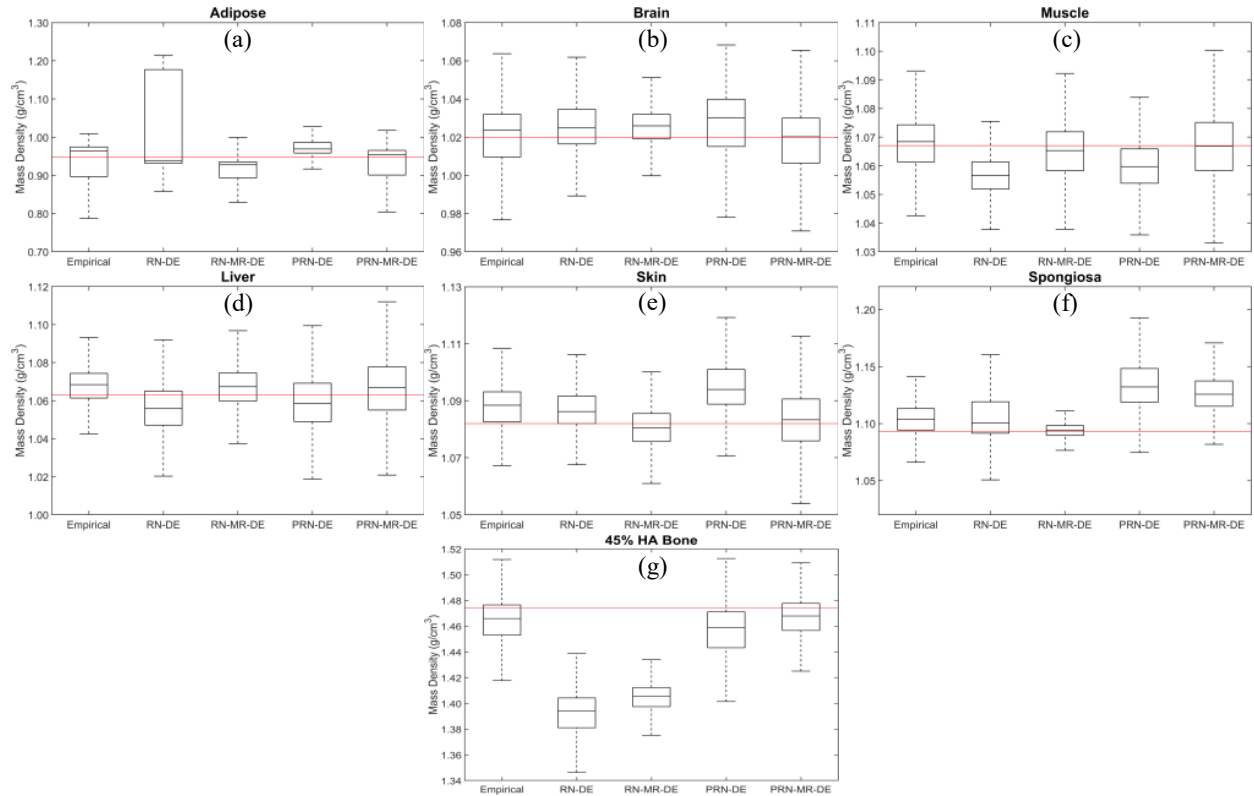

**Figure 3.** Boxplot of mass density distributions generated by the empirical model, RN-DE, RN-MR-DE, PRN-DE, and PRN-MR-DE for tissue substitute phantoms including (a) adipose, (b) brain, (c) muscle, (d) liver, (e) skin, (f) spongiosa, and (g) 45% HA bone. The red line presents the reference mass densities.

### 3.2 Retrospective head-and-neck patient analysis

Figure 4 depicts a retrospective density map prediction using HN patient images. Figure 4(a1)-(a2) show the patient anatomy acquired from MRI and DECT scans. Figure 4(b1)-(b4) illustrate the density maps generated from different models. Figure 4(c1) portrays the line profile from the red line in Figure 4(a1)-(a2) across regions of spine vertebrae. At voxel 22 and 91, PRN-DE estimates the mass density equal to 0.95 g/cm³, while the empirical model and PRN-MR-DE predict the value of 1.04 g/cm³. Based on the MRI image from Figure 4(a2), these two voxels correspond to soft tissue, and a typical mass density of soft tissue should range from 1.04 g/cm³ to 1.07 g/cm³ according to ICRU 44 (ICRU44, 1989). Figure 4(c1) also shows that the densities of cervical vertebrae at voxel 56 are: 1.69 g/cm³ from the empirical model and RN-MR-DE; 1.97 g/cm³ from PRN-DE; 1.80 g/cm³ from PRN-MR-DE. Based on ICRP 70 (ICRP70, 1995), the mass density of hydrated trabecular bone is about 1.87 g/cm³. Figure 4(c2) gives the line profile from the purple line in Figure 4(a1)-(a2). From the MRI image (Figure 4(a2)), the voxel 1-8 and voxel 9-13 correspond to adipose and soft tissue, and average mass densities for voxel over these two regions are: 0.93 g/cm³ and 1.03 g/cm³ from PRN-MR-DE; 0.88 g/cm³ and 0.98 g/cm³ from PRN-DE. The expected mass density of adipose is 0.92 g/cm³ from ICRU 44 (ICRU44, 1989). Figure 4(c2) also shows that the mass densities of the temporal bone at voxel 15 are: 2.33 g/cm³ from PRN-DE; 2.15 g/cm³ from PRN-MR-DE; 2.03 g/cm³ from the empirical model. Based on the literature (Peterson and Dechow, 2003), 99.7% of temporal bone data ranges from 1.63 g/cm³ to 2.22 g/cm³. Both Figure 4(c1)-(c2) display that PRN-MR-DE agrees with the trend of VMI HU profiles.

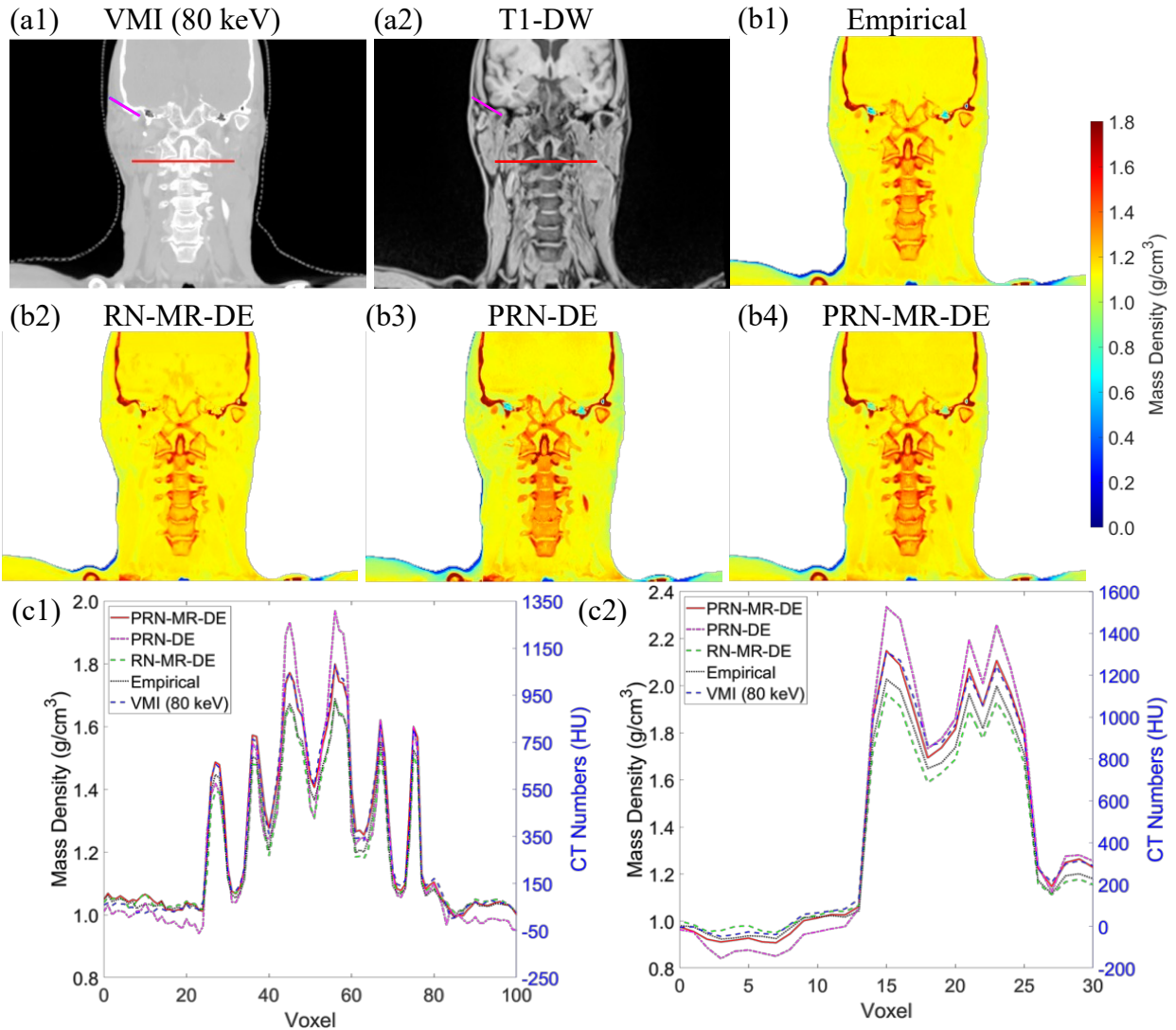

**Figure 4.** Images from (a1) 80-keV VMI and (b1) T1-weighted Dixon water-only (T1-DW) for the same patient. Mass density maps of a patient from an HN site generated by (b1) the empirical model (Eq. (8)), (b2) conventional ResNet trained by MRI and DECT (RN-MR-DE) images, (b3) physics-constrained ResNet trained by DECT (PRN-DE) images, and (b4) physics-constrained ResNet trained by MRI and DECT (PRN-MR-DE) images. (c1) The line profile of the red line from (a1)-(a2). (c2) The line profile of the purple line from (a1)-(a2).

## 4 Discussion

DECT has been clinically evaluated and investigated to derive accurate RSP maps, which can be adapted by commercial treatment planning systems for proton treatment planning (Wohlfahrt *et al.*, 2017; Wohlfahrt *et al.*, 2020). These approaches can be classified into physics-based (Bourque *et al.*, 2014; Möhler *et al.*, 2016; Zhu and Penfold, 2016) and machine learning (ML)-based (Su *et al.*, 2018) dual-energy models, and both approaches aim to support proton analytical dose calculation. The previous ML methods require a separate measurement for phantoms' RSP or mass density values as the ground truth to train ML models such that the measurement uncertainty from calibration phantoms is inherently rooted in the ML models.

We propose a physics-constrained DL-based multimodal imaging (PDMI) framework, for the first time, to embed physics insights into DL training and leverage the information from MRI to deliver accurate patient material mass density maps. The physics-constrained training does not require separate measurement of mass densities for calibration phantoms as the ground truth targets; instead, the targets are CT numbers from VMI that can be consistently acquired from DECT images as the acquisition for DL model inputs. For instance, Table 5 indicates that RN-MR-DE without physics training underestimates the mass densities for adipose and 45% HA bone by 2.86% and 4.28%, compared to PRN-MR-DE with physics training. Additional MRI information can also improve the accuracy of mass density inference for DL models. For example, by training DL models with additional MRI images, conventional RN-MR-DE improve the errors by 4.07% and 1.15% for adipose and spongiosa, compared to RN-DE trained from DECT-only images. Although Table 5 shows that the empirical model achieves the minimum error for adipose, PRN-MR-DE still shows comparable results since the difference is 0.1%. Figure 3 depicts that PRN-MR-DE predicts the closest median values of mass density for most phantoms. PRN-MR-DE generally can reach the optimal error values for most tissue substitute phantoms.

For the retrospective patient analysis, the line profiles in Figure 4(c1)-(c2) show that the mass density curves by PRN-MR-DE agree with the trend of 80-keV VMI, and the density values for adipose, soft tissue, and bone are within a reasonable range published in literature (ICRU44, 1989; ICRP70, 1995; Peterson and Dechow, 2003). However, Figure 4(c1) shows that the empirical model tends to underestimate the mass density for cervical vertebrae. Figure 4(c2) displays that PRN-DE underrates the mass density for adipose and soft tissue and results in temporal density variations larger than the range found in the literature (Peterson and Dechow, 2003). These observations indicate that the additional MRI information can increase the accuracy of DL models. Meanwhile, RN-MR-DE without physics constraints could underestimate temporal density in some regions. For example, Figure 4(c2) depicts that RN-MR-DE predicts the temporal density of 1.59 g/cm$^3$ at voxel 18, outside 99.7% of the temporal density data (Peterson and Dechow, 2003). ResNet trained without MRI images and physics insights exhibited similar patterns between the phantom and retrospective patient studies. Both models could underestimate the mass densities for low- and high-density materials.

Hünemohr et al. (Hünemohr et al., 2014) concluded that the uncertainty of RSP was dominated by mass density prediction. Proton Monte Carlo dose calculation has been recommended for treatment planning when patient images involve heterogeneous tissues (Schuemann et al., 2015; Huang et al., 2018) or surgical implants (Lin et al., 2021). Consequently, we aim to use physics insights to regularize DL models and combine MRI and DECT data to derive accurate mass density maps. The proposed physics-constrained method is based on image domain training, and model accuracy can be compromised by image noise, artifacts, or calibration phantom inhomogeneities. The accuracy of the empirical model given by Eq. (8) is also impacted by the noise of relative electron density maps. Table 5 indicates that RN-MR-DE can deliver minimal spongiosa error due to the phantom inhomogeneity. In contrast, conventional DL model training corresponds each input image voxel to a constant material mass density so that conventional DL models can demonstrate effective noise suppression capability.

Model inference from data belongs to inverse problems, which are ill-posed (O'Sullivan, 1986), and the physics-constrained loss can potentially regularize DL models to remedy this ill-posedness. However, correlating MRI and DECT images to material mass density maps is subject to high dimensional data learning that requires certain degrees of extrapolation (Balestriero et al., 2021). Additional biological equivalent and MRI-visible tissue substitute phantoms are essential to extend training data coverage and increase the robustness of DL models. Furthermore, MRI signal intensity is dependent upon T1 and T2 relaxation time constants of the constituent hydrogen nuclei, or spins, within the water molecules in these calibration phantoms, which themselves vary based on molecular motion, size, and interactions between other spins and the surrounding lattice. For instance, cortical bone has a long T1 because water molecules are tightly bound to rigid HA crystals, which does not allow for motion near the Larmor frequency required for efficient T1 relaxation of a spin back to its surrounding lattice, and a very short T2 because there is efficient spin-spin relaxation among the tightly bound water molecules; these effects typically make cortical bone dark on T1- and T2-weighted MR images. While most of the tissue substitute phantoms demonstrate appropriate contrast on the T1-DW, T1-DF, and T2-STIR MR images, the 45% HA phantom does not display the same contrast as cortical bone. Although the 45% HA phantom is also dark on T2-STIR imaging, it is uncharacteristically bright on T1-DW imaging. This is likely due to differences in the micro-structure of the HA phantom and *in vivo* cortical bone, which result in the water molecules in the phantom not as tightly bound to HA crystals as in bone. In the future, the use of ultrashort echo time (UTE) or zero echo time (ZTE) MR sequences could be incorporated to better-delineate cortical bone (Ma *et al.*, 2020). Meanwhile, customized MRI-visible phantoms usually are made with water, fat, or other organic matter, and this approach can introduce additional uncertainty for phantom mass densities from volumetric and weight measurement and phantom preservation. Physics-constrained training has the potential to be a promising tool to eliminate this uncertainty. Future investigation will likewise integrate the proposed PDMI framework into treatment planning systems to evaluate dosimetry impacts for treatment planning with mass density maps generated from different models.

## 5    Conclusions

A PDMI framework was developed to demonstrate the feasibility of using MRI to improve patient mass density maps generated by DECT-only DL methods using tissue substitute surrogates. Physics-constrained training can regularize DL models and enhance model efficacy, especially when the measurement of mass densities for calibration phantoms includes significant uncertainty. The retrospective patient density map showed that tissue mass densities obtained from the proposed physics-constrained model were within expected intervals. The proposed framework has the potential to improve the quality of treatment planning for proton therapy through accurate material mass density maps derived from MRI and DECT images.

**Acknowledgments**

This research is supported in part by the National Cancer Institute of the National Institutes of Health under Award Number R01CA215718 and by the NIBIB of the National Institutes of Health under Award Number R21EB026086.

# References

Balestriero R, Pesenti J and LeCun Y 2021 Learning in High Dimension Always Amounts to Extrapolation *arXiv preprint arXiv:2110.09485*

Bär E, Lalonde A, Royle G, Lu H-M and Bouchard H 2017 The potential of dual-energy CT to reduce proton beam range uncertainties *Medical Physics* **44** 2332-44

Bär E, Lalonde A, Zhang R, Jee K-W, Yang K, Sharp G, Liu B, Royle G, Bouchard H and Lu H-M 2018 Experimental validation of two dual-energy CT methods for proton therapy using heterogeneous tissue samples *Medical Physics* **45** 48-59

Baumann B C, Mitra N, Harton J G, Xiao Y, Wojcieszynski A P, Gabriel P E, Zhong H, Geng H, Doucette A, Wei J, O'Dwyer P J, Bekelman J E and Metz J M 2020 Comparative Effectiveness of Proton vs Photon Therapy as Part of Concurrent Chemoradiotherapy for Locally Advanced Cancer *JAMA Oncology* **6** 237-46

Beaulieu L, Carlsson Tedgren Å, Carrier J-F, Davis S D, Mourtada F, Rivard M J, Thomson R M, Verhaegen F, Wareing T A and Williamson J F 2012 Report of the Task Group 186 on model-based dose calculation methods in brachytherapy beyond the TG-43 formalism: Current status and recommendations for clinical implementation *Medical Physics* **39** 6208-36

Bichsel H 1969 Passage of charged particles through matter. California Univ., Berkeley (USA). Dept. of Physics)

Blumer A, Ehrenfeucht A, Haussler D and Warmuth M K 1987 Occam's Razor *Information Processing Letters* **24** 377-80

Bourque A E, Carrier J-F and Bouchard H 2014 A stoichiometric calibration method for dual energy computed tomography *Physics in Medicine and Biology* **59** 2059-88

Champion K, Lusch B, Kutz J N and Brunton S L 2019 Data-driven discovery of coordinates and governing equations *Proceedings of the National Academy of Sciences* **116** 22445

Chang C-W and Dinh N T 2019 Classification of machine learning frameworks for data-driven thermal fluid models *International Journal of Thermal Sciences* **135** 559-79

Chang C-W, Gao Y, Wang T, Lei Y, Wang Q, Pan S, Sudhyadhom A, Bradley J D, Liu T, Lin L, Zhou J and Yang X 2022 Dual-energy CT based mass density and relative stopping power estimation for proton therapy using physics-informed deep learning *Physics in Medicine & Biology* **67** 115010

Chang C-W, Huang S, Harms J, Zhou J, Zhang R, Dhabaan A, Slopsema R, Kang M, Liu T, McDonald M, Langen K and Lin L 2020 A standardized commissioning framework of Monte Carlo dose calculation algorithms for proton pencil beam scanning treatment planning systems *Medical Physics* **47** 1545-57

Dinges E, Felderman N, McGuire S, Gross B, Bhatia S, Mott S, Buatti J and Wang D 2015 Bone marrow sparing in intensity modulated proton therapy for cervical cancer: Efficacy and robustness under range and setup uncertainties *Radiotherapy and Oncology* **115** 373-8

Domingos P 1999 The Role of Occam's Razor in Knowledge Discovery *Data Mining and Knowledge Discovery* **3** 409-25

Edmund J M, Kjer H M, Van Leemput K, Hansen R H, Andersen J A L and Andreasen D 2014 A voxel-based investigation for MRI-only radiotherapy of the brain using ultra short echo times *Physics in Medicine and Biology* **59** 7501-19

Goitein M 1985 Calculation of the uncertainty in the dose delivered during radiation therapy *Medical Physics* **12** 608-12

Grant K L, Flohr T G, Krauss B, Sedlmair M, Thomas C and Schmidt B 2014 Assessment of an Advanced Image-Based Technique to Calculate Virtual Monoenergetic Computed Tomographic Images From a Dual-Energy Examination to Improve Contrast-To-Noise Ratio in Examinations Using Iodinated Contrast Media *Investigative Radiology* **49**

He K, Zhang X, Ren S and Sun J 2016 Deep Residual Learning for Image Recognition *2016 IEEE Conference on Computer Vision and Pattern Recognition (CVPR)* 770-8

Hornik K, Stinchcombe M and White H 1989 Multilayer feedforward networks are universal approximators *Neural Networks* **2** 359-66

Huang S, Souris K, Li S, Kang M, Barragan Montero A M, Janssens G, Lin A, Garver E, Ainsley C, Taylor P, Xiao Y and Lin L 2018 Validation and application of a fast Monte Carlo algorithm for assessing the clinical impact of approximations in analytical dose calculations for pencil beam scanning proton therapy *Medical Physics* **45** 5631-42

Hünemohr N, Paganetti H, Greilich S, Jäkel O and Seco J 2014 Tissue decomposition from dual energy CT data for MC based dose calculation in particle therapy *Medical Physics* **41** 061714

ICRP70 1995 Basic Anatomical & Physiological Data for use in Radiological Protection - The Skeleton *ICRP Publication 70*

ICRU37 1987 Stopping Powers for Electrons and Positrons *ICRU Publication 37*

ICRU44 1989 Tissue Substitutes in Radiation Dosimetry and Measurement *ICRU Publication 44*

Karniadakis G E, Kevrekidis I G, Lu L, Perdikaris P, Wang S and Yang L 2021 Physics-informed machine learning *Nature Reviews Physics* **3** 422-40

Koivula L, Wee L and Korhonen J 2016 Feasibility of MRI-only treatment planning for proton therapy in brain and prostate cancers: Dose calculation accuracy in substitute CT images *Medical Physics* **43** 4634-42

LeCun Y, Bengio Y and Hinton G 2015 Deep learning *Nature* **521** 436-44

Liang X, Li Z, Zheng D, Bradley J A, Rutenberg M and Mendenhall N 2019 A comprehensive dosimetric study of Monte Carlo and pencil-beam algorithms on intensity-modulated proton therapy for breast cancer *Journal of Applied Clinical Medical Physics* **20** 128-36

Lin L, Taylor P A, Shen J, Saini J, Kang M, Simone C B, II, Bradley J D, Li Z and Xiao Y 2021 NRG Oncology Survey of Monte Carlo Dose Calculation Use in US Proton Therapy Centers *International Journal of Particle Therapy* **8** 73-81

Ma Y-J, Jerban S, Jang H, Chang D, Chang E Y and Du J 2020 Quantitative Ultrashort Echo Time (UTE) Magnetic Resonance Imaging of Bone: An Update *Frontiers in Endocrinology* **11**

Mayneord W 1937 The significance of the roentgen *Acta Int Union Against Cancer* **2** 271-82

McCollough C H, Leng S, Yu L and Fletcher J G 2015 Dual- and Multi-Energy CT: Principles, Technical Approaches, and Clinical Applications *Radiology* **276** 637-53

Medrano M, Liu R, Zhao T, Webb T, Politte D G, Whiting B R, Liao R, Ge T, Porras-Chaverri M A, O'Sullivan J A and Williamson J F 2022 Towards subpercentage uncertainty proton stopping-power mapping via dual-energy CT: Direct experimental validation and uncertainty analysis of a statistical iterative image reconstruction method *Medical Physics* **49** 1599-618

Möhler C, Wohlfahrt P, Richter C and Greilich S 2016 Range prediction for tissue mixtures based on dual-energy CT *Physics in Medicine and Biology* **61** N268-N75

Nair V and Hinton G E 2010 Rectified linear units improve restricted boltzmann machines. In: *Proceedings of the 27th International Conference on International Conference on Machine Learning,* (Haifa, Israel: Omnipress) pp 807–14

O'Sullivan F 1986 A Statistical Perspective on Ill-Posed Inverse Problems *Statistical Science* **1** 502-18

Paganetti H 2012 Range uncertainties in proton therapy and the role of Monte Carlo simulations *Physics in Medicine and Biology* **57** R99-R117

Paganetti H, Jiang H, Parodi K, Slopsema R and Engelsman M 2008 Clinical implementation of full Monte Carlo dose calculation in proton beam therapy *Physics in Medicine and Biology* **53** 4825-53

Paszke A, Gross S, Massa F, Lerer A, Bradbury J, Chanan G, Killeen T, Lin Z, Gimelshein N, Antiga L, Desmaison A, Kopf A, Yang E, DeVito Z, Raison M, Tejani A, Chilamkurthy S, Steiner B, Fang L, Bai J and Chintala S 2019 PyTorch: An Imperative Style, High-Performance Deep Learning Library *Advances in Neural Information Processing Systems* **32**

Patel B N, Thomas J V, Lockhart M E, Berland L L and Morgan D E 2013 Single-source dual-energy spectral multidetector CT of pancreatic adenocarcinoma: Optimization of energy level viewing significantly increases lesion contrast *Clinical Radiology* **68** 148-54

Peterson J and Dechow P C 2003 Material properties of the human cranial vault and zygoma *The Anatomical Record Part A: Discoveries in Molecular, Cellular, and Evolutionary Biology* **274A** 785-97

Rank C M, Hünemohr N, Nagel A M, Röthke M C, Jäkel O and Greilich S 2013 MRI-based simulation of treatment plans for ion radiotherapy in the brain region *Radiotherapy and Oncology* **109** 414-8

Ratner B 2011 *Statistical and Machine-Learning Data Mining: Techniques for Better Predictive Modeling and Analysis of Big Data*: CRC Press)

Rutherford R A, Pullan B R and Isherwood I 1976 Measurement of effective atomic number and electron density using an EMI scanner *Neuroradiology* **11** 15-21

Saini J, Maes D, Egan A, Bowen S R, St James S, Janson M, Wong T and Bloch C 2017 Dosimetric evaluation of a commercial proton spot scanning Monte-Carlo dose algorithm: comparisons against measurements and simulations *Physics in Medicine & Biology* **62** 7659-81

Schneider U, Pedroni E and Lomax A 1996 The calibration of CT Hounsfield units for radiotherapy treatment planning *Physics in Medicine and Biology* **41** 111-24

Schneider W, Bortfeld T and Schlegel W 2000 Correlation between CT numbers and tissue parameters needed for Monte Carlo simulations of clinical dose distributions *Physics in Medicine and Biology* **45** 459-78

Scholey J E, Chandramohan D, Naren T, Liu W, Larson P E Z and Sudhyadhom A 2021 Technical Note: A methodology for improved accuracy in stopping power estimation using MRI and CT *Medical Physics* **48** 342-53

Scholey J E, Vinas L, Kearney V P, Yom S S, Larson P E Z, Descovich M and Sudhyadhom A 2022 Improved accuracy of relative electron density and proton stopping power ratio through CycleGAN machine learning *Physics in Medicine & Biology*

Schuemann J, Giantsoudi D, Grassberger C, Moteabbed M, Min C H and Paganetti H 2015 Assessing the Clinical Impact of Approximations in Analytical Dose Calculations for Proton Therapy *Int J Radiat Oncol Biol Phys* **92** 1157-64

Spiers F W 1946 Effective Atomic Number and Energy Absorption in Tissues *The British Journal of Radiology* **19** 52-63

Su K-H, Kuo J-W, Jordan D W, Van Hedent S, Klahr P, Wei Z, Al Helo R, Liang F, Qian P, Pereira G C, Rassouli N, Gilkeson R C, Traughber B J, Cheng C-W and Muzic R F 2018 Machine learning-based dual-energy CT parametric mapping *Physics in Medicine & Biology* **63** 125001

Sudhyadhom A 2017 Determination of mean ionization potential using magnetic resonance imaging for the reduction of proton beam range uncertainties: theory and application *Physics in Medicine & Biology* **62** 8521-35

Wang T, Ghavidel B B, Beitler J J, Tang X, Lei Y, Curran W J, Liu T and Yang X 2019 Optimal virtual monoenergetic image in "TwinBeam" dual-energy CT for organs-at-risk delineation based on contrast-noise-ratio in head-and-neck radiotherapy *Journal of Applied Clinical Medical Physics* **20** 121-8

Wohlfahrt P, Möhler C, Enghardt W, Krause M, Kunath D, Menkel S, Troost E G C, Greilich S and Richter C 2020 Refinement of the Hounsfield look-up table by retrospective application of patient-specific direct proton stopping-power prediction from dual-energy CT *Medical Physics* **47** 1796-806

Wohlfahrt P, Möhler C, Hietschold V, Menkel S, Greilich S, Krause M, Baumann M, Enghardt W and Richter C 2017 Clinical Implementation of Dual-energy CT for Proton Treatment Planning on Pseudo-monoenergetic CT scans *International Journal of Radiation Oncology*Biology*Physics* **97** 427-34

Wohlfahrt P, Möhler C, Troost E G C, Greilich S and Richter C 2019 Dual-Energy Computed Tomography to Assess Intra- and Inter-Patient Tissue Variability for Proton Treatment Planning of Patients With Brain Tumor *Int J Radiat Oncol Biol Phys* **105** 504-13

Xie Y, Ainsley C, Yin L, Zou W, McDonough J, Solberg T D, Lin A and Teo B-K K 2018 Ex vivo validation of a stoichiometric dual energy CT proton stopping power ratio calibration *Physics in Medicine & Biology* **63** 055016

Yang M, Virshup G, Clayton J, Zhu X R, Mohan R and Dong L 2010 Theoretical variance analysis of single- and dual-energy computed tomography methods for calculating proton stopping power ratios of biological tissues *Physics in Medicine and Biology* **55** 1343-62

Yepes P, Adair A, Grosshans D, Mirkovic D, Poenisch F, Titt U, Wang Q and Mohan R 2018 Comparison of Monte Carlo and analytical dose computations for intensity modulated proton therapy *Physics in Medicine & Biology* **63** 045003

Yu L, Christner J A, Leng S, Wang J, Fletcher J G and McCollough C H 2011 Virtual monochromatic imaging in dual-source dual-energy CT: Radiation dose and image quality *Medical Physics* **38** 6371-9

Yu L, Leng S and McCollough C H 2012 Dual-Energy CT–Based Monochromatic Imaging *American Journal of Roentgenology* **199** S9-S15

Zhu J and Penfold S N 2016 Dosimetric comparison of stopping power calibration with dual-energy CT and single-energy CT in proton therapy treatment planning *Medical Physics* **43** 2845-54